\documentclass[twocolumn,aps,showpacs,floatfix,prc]{revtex4}
\usepackage[dvips]{epsfig}
\newcommand{\etal}{{\it et al.}}
\begin{document}
\title{Sequential melting of charmonium states in an expanding Quark Gluon Plasma and $J/\psi$ suppression at RHIC and LHC energy collisions}
\author{A. K. Chaudhuri}
\email[E-mail:]{akc@veccal.ernet.in}
\affiliation{Variable Energy Cyclotron Centre,\\ 1/AF, Bidhan Nagar,
Kolkata 700~064, India}

\begin{abstract}

We have developed a hydrodynamic model to study sequential melting of charmonium states in an expanding QGP medium.   According to the initial fluid temperature profile, $J/\psi$'s are randomly distributed in the transverse plane. As the fluid evolve in time, the free streaming $J/\psi$'s are suppressed if the local fluid  temperature exceeds a critical temperature.
PHENIX data on the centrality dependence of $J/\psi$ suppression in Au+Au collisions at mid-rapidity are explained by sequential melting of the charmonium states, $\chi_c$, $\psi\prime$ and $J/\psi$, in the expanding medium. The critical temperatures 
$T_{J/\psi} \approx2.09T_c$ and $T_\chi=T_{\chi_c}=T_{\psi\prime} \approx 1.1T_c$ agree with lattice motivated calculations. The feed-down fraction $F$ depend on whether the cold nuclear matter effect is included or not. It changes from $F=0.3$ with cold nuclear matter effect included to $F=0.5$ when the effect is neglected.   Model fails to reproduce the PHENIX data
on the centrality dependence of
$J/\psi$ suppression in Cu+Cu collisions at mid-rapidity, indicating that the mechanism of $J/\psi$ suppression is different in Au+Au and in  Cu+Cu collisions.  
We also use the model to predict for the centrality dependence of $J/\psi$ suppression in Pb+Pb collisions at LHC energy, $\sqrt{s}$=5500 GeV. In LHC energy, $J/\psi$'s are more suppressed in mid central collisions than in Au+Au collisions at RHIC energy.
\end{abstract}  

\maketitle
 
\section{introduction} 
 
In  relativistic  heavy  ion  collisions $J/\psi$ suppression has
been recognized as an important tool  to  identify  the  possible
phase transition to quark-gluon plasma. Because of the large mass
of  the  charm  quarks,  $c\bar{c}$ pairs are produced on a short
time scale. Their tight binding also makes them 'nearly' immune  to  final
state interactions. Their evolution probes the state of matter in
the  early  stage  of the collisions. Matsui and Satz  \cite{Matsui:1986dk}, 
predicted that in presence of quark-gluon plasma  (QGP),  binding
of a $c\bar{c}$  pair  into  a  $J/\psi$  meson will be hindered,
leading to the  so  called  $J/\psi$  suppression  in  heavy  ion
collisions  \cite{Matsui:1986dk} .  Over  the  years,  several  groups have
measured the $J/\psi$ yield in heavy ion collisions (for a review
of the data prior to RHIC energy collisions, and the interpretations see Refs.  \cite{vo99,ge99}).
In  brief,  experimental  data do show suppression. However, this
could be attributed to the conventional nuclear absorption,  also
present in $pA$ collisions.

PHENIX  collaboration  has undertaken the task to characterise medium effect on $J/\psi$ production in nuclear collisions. 
They have measured $J/\psi$ yield in p+p collisions at RHIC and obtained the  reference for basic invariant yield \cite{Adler:2003qs,Adler:2005ph,Adare:2006kf}. Measurements of $J/\psi$ production  in d+Au collisions
\cite{Adler:2005ph,Adare:2007gn} give reference for cold nuclear matter effects. $J/\psi$ production in d+Au collisions are consistent with cold nuclear matter effect quantified in a Glauber model of nuclear absorption with $\sigma_{abs}=2\pm 1$ mb \cite{Vogt:2005ia}.
Cold and hot nuclear matter effects are studied  in
Au+Au and Cu+Cu collisions, where yields are measured as a function of collision centrality \cite{Adler:2003rc,Adare:2006ns,Adare:2008sh}. 
In Au+Au \cite{Adler:2003rc,Adare:2006ns} and Cu+Cu \cite{Adare:2008sh} collisions, data are taken
at mid-rapidity ($|y| < .35$) and at forward rapidity ($1.2<y<2.2$). $J/\psi$'s  are more suppressed at forward rapidity than at mid rapidity. It was also noted that 
at comparable participant number,   $J/\psi$'s are suppressed similarly in Au+Au and in Cu+Cu collisions \cite{Adare:2008sh}.

At RHIC energy, it has been
argued that rather than suppression, charmonium's will be enhanced
\cite{ Thews:2000rj,Braun-Munzinger:2000px}. 
Due to large initial energy, large number of $c\bar{c}$ pairs will be
produced in initial hard scatterings. Recombination of $c\bar{c}$
can occur enhancing the charmonium production.
Both the PHENIX data on
$J/\psi$ production in Au+Au and in Cu+Cu collisions, are not consistent
with models which predict $J/\psi$ enhancement
\cite{ Thews:2000rj,Braun-Munzinger:2000px}.  
The cold nuclear matter effect, quantified by the Glauber model of nuclear absorption with $J/\psi$-nucleon absorption cross-section $\sigma_{abs}=2\pm 1$ mb,    is consistent only with  peripheral Cu+Cu collisions. 
In all the centrality ranges of Au+Au collision, 
or in central Cu+Cu collisions, suppression is beyond the cold nuclear matter effect.

Blaizot \etal \cite{Blaizot:2000ev,Blaizot:1996nq} proposed a phenomenological model, called the threshold model,
to describe $J/\psi$ suppression in Pb+Pb collisions at SPS energy.
In the threshold model, to mimic the onset of deconfining phase transition above a critical energy density and subsequent melting of $J/\psi$, $J/\psi$ suppression is linked with the local transverse density. If the local transverse density at the point where $J/\psi$ is formed exceed a critical or threshold value, $J/\psi$'s are melted.
Recently, 
in the threshold model, we have analysed the PHENIX data on the centrality dependence of $J/\psi$ suppression in Au+Au/Cu+Cu collisions at mid-rapidity \cite{Chaudhuri:2006fe,Chaudhuri:2007qz,Chaudhuri:2007yy}. Sequential melting of charmonium states,
$\chi_c$, $\psi^\prime$ and $J/\psi$ above threshold density,
$n_{\chi_c}=n_{\psi\prime}=n_\chi$ and $n_{J/\psi}$ explains the PHENIX data on $J/\psi$ suppression in mid-rapidity Au+Au collisions \cite{Chaudhuri:2007yy}.
It was also observed that the feed-down  fraction $F$, from higher states $\chi_c$ and $\psi^\prime$, depends on the quantum of nuclear absorption. Equivalent fit to the data could be obtained by increasing the fraction F and decreasing the $J/\psi$-nucleon absorption cross-section or the vice-versa.  The threshold model 
ignore the expansion of the medium. Also, in the threshold model, while it is assumed that the threshold density is proportional to the critical energy density above with
charmonium states melt, the exact relation between them is rather obscure. 

Recently  Gunji \etal \cite{Gunji:2007uy} analysed the PHENIX mid-rapidity data on the centrality dependence of $J/\psi$ suppression in Au+Au collisions . They developed a "Hydro+$J/\psi$" model. The QGP fluid evolves in 3+1 dimensions. At the initial time, in accordance to the fluid temperature, $J/\psi$'s are randomly distributed in the fluid. As the fluid evolve, free streaming $J/\psi$'s are melted if the local fluid temperature exceeds a critical value. 
The experimental $J/\psi$ suppression pattern in mid-rapidity Au+Au collisions is well explained by sequential melting of $\chi_c$, $\psi\prime$ and $J/\psi$ in the dynamically expanding fluid. 
The estimated melting temperatures, $T_{J/\psi}=2.02 T_c$,
$T_{\chi_c}=T_{\psi\prime}=1.22 T_c$   are in agreement with the lattice motivated calculations \cite{Satz:2006kb}. The fraction of the higher states $(\chi_c+\psi^\prime)$ is estimated to be $F=0.3$. 
It may be mentioned that experimentally, the feed-down fraction $F$ is largely uncertain \cite{Abt:2002vq}.  
Measurements are available over a wide range of energy $\sqrt{s}$=8.5-1800 GeV. Measured values show considerable variation, $F=0.15-0.74$. The estimated fraction $F$=0.3 is well within the largely uncertain range of measurements. Incidentally, measurement  at an energy comparable to RHIC energy $\sqrt{s}$=200 GeV is not available.

In the present paper, we have developed a hydrodynamic model for $J/\psi$ suppression in heavy ion collisions. The aim was to verify the results obtained by Gunji \etal \cite{Gunji:2007uy} and
to use the model to predict for the suppression pattern at Pb+Pb collisions at LHC energy.
The model is similar to that of Gunji \etal \cite{Gunji:2007uy}, though there are some differences in details.
Gunji \etal \cite{Gunji:2007uy} solved hydrodynamic equations in 3+1 dimensions. The initial fluid energy density in the transverse plane was parameterised in proportion to the Glauber model calculation of hard collisions. We have  
solved the hydrodynamic equations in a 2+1 dimensions assuming longitudinal boost-invariance. The model is thus limited to $J/\psi$ production at mid-rapidity only. The initial fluid energy density in the transverse plane is assumed to be proportional to the Glauber model calculation with 75\% soft collisions and 25\% hard collisions \cite{QGP3}. 
As it will be shown later, PHENIX data \cite{Adler:2003rc,Adare:2006ns} on the centrality dependence of $J/\psi$ suppression in Au+Au collisions are reproduced in the model with melting temperatures and feed-down fraction close to the values obtained by Gunji \etal \cite{Gunji:2007uy}.   We have also analysed the recently published Cu+Cu data \cite{Adare:2008sh}. Hydro+$J/\psi$ model with melting temperatures extracted from the analysis of Au+Au data, is not consistent with the $J/\psi$ suppression in Cu+Cu collisions. The model produces more suppression than required by the data.
  Clearly, $J/\psi$ suppression mechanism in Au+Au and Cu+Cu collisions is not identical. 
We also use the model to predict for the centrality dependence of $J/\psi$ suppression in Pb+Pb collisions at LHC energy ($\sqrt{s}$=5500 GeV). In LHC energy collisions, $J/\psi$'s are more suppressed than in Au+Au collisions at RHIC.
 
The plan of the paper is as follows: in section II, we briefly describe the hydrodynamic model. The explicit mechanism followed to suppress $J/\psi$'s in the expanding fluid is also discussed in section II. In section III, we have analysed  PHENIX data on the centrality dependence of $J/\psi$ suppression in Au+Au collisions and obtain the parameters of the model. We also analyse the PHENIX data on the centrality dependence of $J/\psi$ suppression in Cu+Cu collisions. 
In section IV we give predictions for $J/\psi$ suppression in Pb+Pb collisions at LHC
energy collisions. Lastly.
summary and conclusions are drawn in section V.

\section{The hydrodynamic model for $J/\psi$ suppression}

\subsection{Hydrodynamic model for QGP evolution}

Details of the hydrodynamic model used here can be 
found in \cite{QGP3}.  In \cite{QGP3}, Kolb and Heinz, assuming longitudinal boost-invariance, solved the energy-momentum conservation equation $\partial_\mu T^{\mu\nu}=0$, in 2+1 dimensions. Hydrodynamic models require   energy density, fluid velocity distributions at the initial time $\tau_i$. In \cite{QGP3},
the initial energy density of the fluid in  the transverse plane was parameterised as, 

\begin{equation} \label{eq1}
\varepsilon({\bf b},x,y) =\varepsilon_0 [0.75 N_{part}({\bf b},x,y) +.25 N_{coll}({\bf b},x,y)]
\end{equation}

\noindent where $N_{part}({\bf b},x,y)$ and $N_{coll}({\bf b},x,y)$ are the transverse profile for the participant number and binary collisions number in an  impact parameter {\bf b}  Au+Au collision. The initial fluid velocity was assumed to be zero, $v_x({\bf b},x,y)=v_y({\bf b},x,y)=0$. The constant $\varepsilon_0$ depend only on the collision energy and not on centrality of the collisions. The initial time $\tau_i$ and the constant $\varepsilon_0$ chosen to reproduce the $p_T$ distribution of identified particles in central Au+Au collisions. 
 For b=0 Au+Au collisions, it correspond to central energy density, $\varepsilon$=30 $GeV/fm^{3}$ or central entropy density  $S_{ini}$=110 $fm^{-3}$, at the initial time $\tau_i$=0.6 fm/c. 
   Kolb and Heinz \cite{QGP3}, 
used an equation state (EOS-Q) incorporating 1st order phase transition with critical temperature $T_c$=164 MeV. The quark phase was modeled by the bag equation of state for u,d,s quarks and gluons. For the hadronic phase, resonance hadron gas equation of state was used. The bag constant was obtained by using the Maxwell construct at the critical temperature $T_c$=164 MeV.   

\begin{figure}[t]
\includegraphics[bb=30 193 567 769
 ,width=0.9\linewidth,clip]{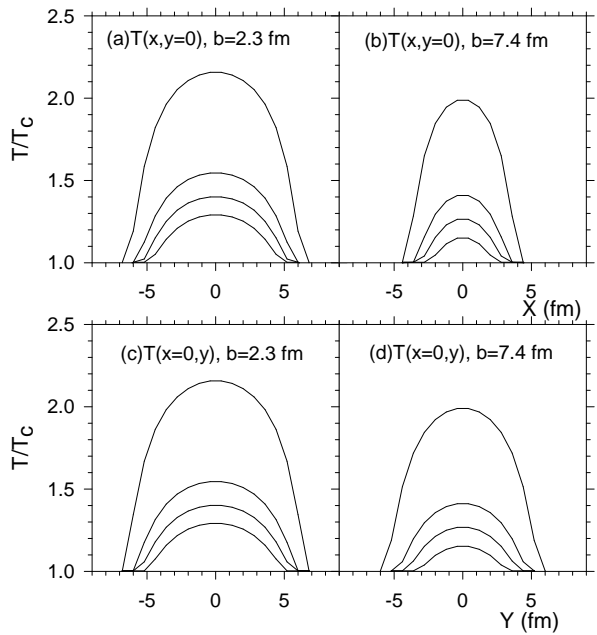}
\caption{In panels (a) and (c), the evolution of local temperature $T(x,y=0)$ as a function of x and $T(x=0,y)$ as a function of y, in b=2.3 fm Au+Au collisions are shown. In the panels (b) and (d) the same are shown for b=7.4 fm Au+Au collisions. In each panel, the different lines, from top to bottom correspond to $\tau_i$=0.6,1.6,2.6 and 3.6 fm.}
 \label{F1}
\end{figure}

\begin{figure}[t]
\includegraphics[bb=44 299 527 767
 ,width=0.9\linewidth,clip]{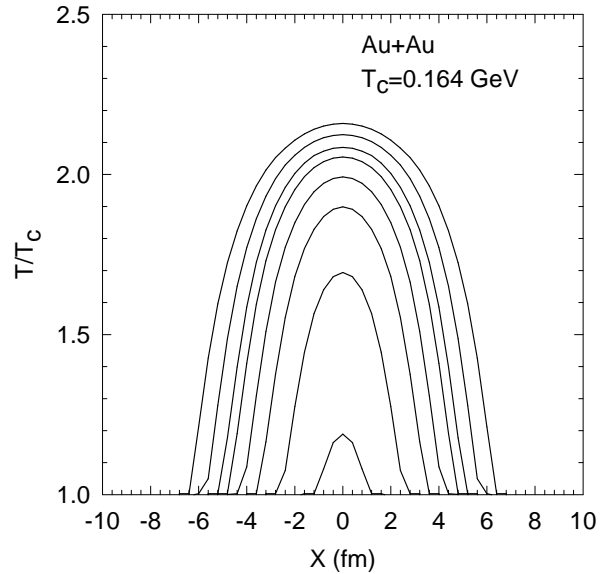}
\caption{Initial temperature, in unit of the critical temperature, of the QGP fluid in Au+Au collisions at impact parameter b=
 2.3, 4.1, 5.2, 6.2, 7.4, 8.7, 10.5 and 13.0 fm (from top to bottom).
They corresponds to  
0-5\%, 5-10\%, 10-15\%, 15-20\%, 20-30\%, 30-40\%, 40-60\% and 60-93\% centrality Au+Au collisions.}
 \label{F2}
\end{figure}

\begin{figure}[t]
\includegraphics[bb=45 294 532 767
 ,width=0.9\linewidth,clip]{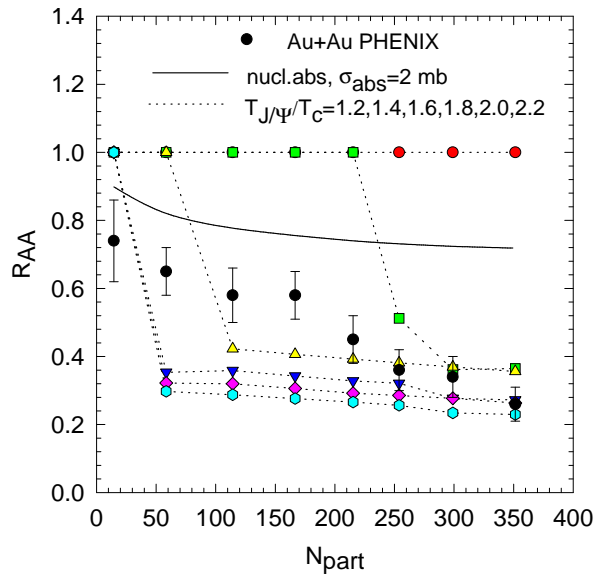}
\caption{(color online) The black filled circles are PHENIX data on the centrality dependence of $J/\psi$ suppression in Au+Au collisions. The black line is the cold nuclear matter effect. computed in a Glauber model with $J/\psi$-nucleon absorption cross-section $\sigma_{abs}$=2 mb. The dotted lines with filled colored symbols are $J/\psi$ survival probability in the present "Hydro+$J/\psi$" model, for melting temperature $T_{J/\psi}/T_c$=1.2,1.4,1.6,1.8,2.0 and 2.2 respectively}
 \label{F3}
\end{figure}

Evolution of QGP fluid in the model is described in detail in \cite{QGP3}. For completeness, in Fig.\ref{F1}(a) and (b), we have shown the
evolution of local temperature $T(x,y=0)$, in unit of the critical temperature $T_c$, in b=2.38 fm (panel (a)) and b=7.4 fm (panel (b)) Au+Au collisions. In Fig.\ref{F1}(c) and (d) local temperature $T(x=0,y)$ as a function of y is shown. The different lines (from top to bottom), in each panel, corresponds to time $\tau_i$=0.6, 1.6, 2.6 and 3.6 fm/c. 
Lattice motivated calculations indicate the just at the critical temperature $T_c$, all the charmonium states are not dissolved \cite{Satz:2006kb}.
The ground state 
$J/\psi(1S)$ can survive in QGP environment up to a temperature $T_{J/\psi}\approx 2.1 T_c$. The excited states $\chi_c (1P)$ and $\psi^\prime (2S)$ on the other hand cannot survive hot QGP.
They are dissolved in  much cooler QGP, $T_{\chi_c}\approx 1.2 T_c$ and $T_{\psi\prime}\approx 1.1 T_c$.
From Fig.\ref{F1}, it is obvious that while the excited states
$\chi_c$ and $\psi\prime$ will be melted in 
{\bf b}=2.3 and 7.4 fm collisions, the ground state can be melted only in {\bf b}=2.3 fm Au+Au collisions. The peak temperature in a b=7.4 fm collision is well below the melting temperature of the ground state $J/\psi$.
Indeed, without any detailed calculations, from the initial peak temperatures in different centrality collisions,
one can very well bound the melting temperature for $J/\psi$ and the states $\chi_c$ and $\psi^\prime$. 
Initial temperature $T(x,y=0)$ of the fluid in b=
2.3, 4.1, 5.2, 6.2, 7.4, 8.7, 10.5 and 13.0 fm Au+Au collisions are shown in Fig.\ref{F2}. Roughly they corresponds to 
0-5\%, 5-10\%, 10-15\%, 15-20\%, 20-30\%, 30-40\%, 40-60\% and 60-93\% Au+Au collisions.  
In 0-5\% centrality collision, peak temperature is only $2.16 T_c$. As the collision centrality decreases, peak temperature decreases and in the most peripheral (60-93\% centrality) collisions peak temperature is only $1.16 T_c$. As shown in Fig.\ref{F3},
in all these centrality ranges of Au+Au collisions $J/\psi$'s are suppressed \cite{Adler:2003rc,Adare:2006ns}.
Presumably, the suppression in 60-93\% centrality collisions is  due to melting of the higher 
states $\chi_c$ and $\psi^\prime$ only. One then obtain an upper bound for the melting temperature of the states $\chi_c$ and $\psi\prime$, $T_{\chi_c},T_{\psi\prime} \leq 1.16 T_c$. Similarly, in 0-5\% centrality collisions, if the ground state $J/\psi$'s are dissolved then  melting temperature can be bounded from above, $T_{J/\psi} \leq 2.16 T_c$. The bounds on the melting temperature  are close to the lattice results \cite{Satz:2006kb} or to the values obtained by Gunji \etal \cite{Gunji:2007uy} from the analysis of PHENIX data.
 
\subsection{$J/\psi$ suppression in the QGP fluid}

To obtain the survival probability of $J/\psi$'s in
an expanding QGP fluid, we proceed as follows:
at the initial time $\tau_i=0.6 fm/c$,  according to the initial spatial distribution of the fluid temperature, we  randomly distribute 
$J/\psi$ in the transverse plane $x-y$.
The transverse momentum of initial $J/\psi$ are distributed according to the power law $A /(1+(p_T/B)^2)^6$, which rather well describe the invariant distribution of measured $J/\psi$'s in p+p collisions \cite{Adler:2003qs,Adler:2005ph,Adare:2006kf}.  The initial $J/\psi$'s are assumed to follow a free streaming path unless dissolved in the medium. To follow the path, to each $J/\psi$, we assigned a random direction vector.
It may be mentioned that while assumption of free streaming do not affect the centrality dependence of $J/\psi$ suppression , it will   affect the $p_T$ distribution \cite{Gunji:2007uy}. Experimentally, $p_T$ distribution of $J/\psi$'s are broadened. The free streaming assumption will not reproduce the $p_T$ broadening of $J/\psi$'s.

\begin{figure}[t]
\includegraphics[bb=45 294 532 767
 ,width=0.9\linewidth,clip]{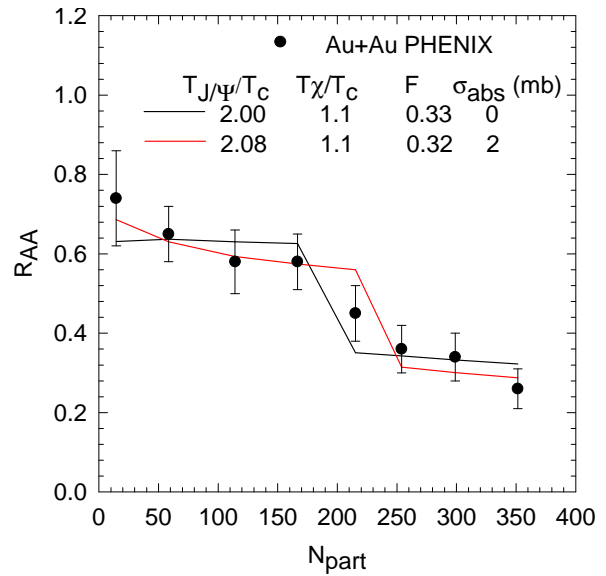}
\caption{(color online) The black filled circles are PHENIX data on the centrality dependence of $J/\psi$ suppression in Au+Au collisions. The black and red lines are best fit to the data in the model, with
parameter values indicated in the figure.}
 \label{F4}
\end{figure}

Following Gunji \etal \cite{Gunji:2007uy}, the survival probability of a $J/\psi$ inside the expanding QGP is defined as,

\begin{equation} \label{eq1a}
S_{J/\psi}(\tau)=exp\left[-\int_{\tau_i}^\tau \Gamma_{dis}(T({\bf x}(\tau^\prime))) d\tau^\prime \right]
\end{equation}

\noindent where $T({\bf x})$ is the temperature of the fluid at the transverse position ${\bf x}$, $\Gamma_{dis}(T)$ is the decay width of $J/\psi$ at temperature $T$. $\tau_i$ is the initial time for hydrodynamic evolution. We continue the evolution till the freeze-out temperature $T_F$=130 MeV.
For the decay width of $J/\psi$, we made the simple choice, 

\begin{eqnarray} \nonumber
\Gamma_{dis}(T) &= & \infty; \hspace{1cm} T>T_{J/\psi} \\
\Gamma_{dis}(T) &= &0; \hspace{1cm}T<T_{J/\psi} \label{eq2}
\end{eqnarray}

In Eq.\ref{eq2}, decay width of $J/\psi$   abruptly changes from $\infty$ to 0 at the melting temperature $T_{J/\psi}$. It neglects 
  the broadening of charmonium states below the critical temperature $T_c$.  In a hot pion gas widths of decay and dissociation channels  of the charmonium states  $\psi\prime$, $\chi_c$ and $J/\psi$ to $D\bar{D}$,
$D^*\bar{D}$, $D\bar{D^*}$, $D^* \bar{D}^*$ pairs are enhanced \cite{Fuchs:2004fh}.  
Lattice QCD calculations \cite{Asakawa:2003re} also indicate D- and B-like states can exist in sQGP. They can provide resonant cross section for heavy quarks. Recently van Hees \etal \cite{van Hees:2005wb}, in a relativistic Langevin approach showed that resonant interactions play an essential part in thermalisation and collective flow of charm and bottom quarks. D- and B-like states in sQGP can also have direct impact on $J/\psi$ production, facilitating regeneration. Survival probability  (Eq.\ref{eq1a}) also neglects regeneration of charmonium states in sQGP. 

Any model of $J/\psi$ suppression must account for the experimental observation that a substantial fraction of the measured $J/\psi$'s are from decay of the excited charmonium states $\chi_c$ and $\psi\prime$ \cite{Abt:2002vq}. 
To calculate the survival probability of the excited states $\chi_c$ and $\psi^\prime$, we use the same procedure as described above
for the ground state $J/{\psi}$. 
Above $T_{\chi_c}$ and $T_{\psi\prime}$, the excited states
$\chi_c$ and $\psi^\prime$ are assumed to melt. Further noting that the lattice motivated calculation indicate $T_{\chi_c} \approx T_{\psi^\prime}$, we define a
common temperature $T_\chi=T_{\chi_c}=T_{\psi\prime}$, above which all the states $\chi_c$ and $\psi^\prime$ are dissolved. For feed-down fraction $F$, the $J/\psi$ survival probability is then obtained as,

\begin{equation}
S_{QGP}=(1-F) S_{J/\psi} + F S_\chi
\end{equation}

As mentioned earlier, PHENIX collaboration in d+Au collisions has studied cold nuclear matter effect on $J/\psi$ suppression
 \cite{Adler:2005ph,Adare:2007gn}. $J/\psi$'s are suppressed in d+Au collisions also. The suppression is consistent with Glauber model of nuclear absorption with $J/\psi$-nucleon absorption cross-section $\sigma_{abs} =2\pm 1$ mb. If cold nuclear matter effect is taken into account, the survival probability of $J/\psi$ can be obtained as,

\begin{equation}
S_{J/\psi}= S_{QGP}\times S_{CNM},
\end{equation}

\noindent where $S_{CNM}$ is the survival probability in cold nuclear matter calculated in a Glauber model.

\section{RESULTS}

\subsection{$J/\psi$ suppression in Au+Au collisions}

PHENIX data \cite{Adler:2003rc,Adare:2006ns} on the centrality dependence of $J/\psi$ suppression are shown in Fig.\ref{F3}. 
In all the centrality ranges of collisions, $J/\psi$'s are suppressed, suppression increasing with collision centrality. One also notes that around $N_{part}\approx$150, there is a distinct change of slope in the suppression. 
The black solid line in Fig.\ref{F3} is an estimate of cold nuclear matter effect, calculated in the Glauber model with $\sigma_{abs}$=2 mb. Experimentally, $J/\psi$'s are more suppressed than in the Glauber model calculation. Evidently, data demand suppression in addition to the nuclear absorption.
In Fig.\ref{F3}, centrality dependence of $J/\psi$ suppression in the present "Hydro+$J/\psi$" model,
for a choice of melting  temperatures $T_{J/\psi}/T_c$=1.2,1.4,1.6,1.8,2.0 and 2.2,
 are shown.  As expected, for $T_{J/\psi}=2.2T_c$, $J/\psi$ are not suppressed in the QGP medium, the fluid temperature is below the melting temperature. $J/\psi$'s are increasingly suppressed as the melting temperature is lowered.  
It is apparent from Fig.\ref{F3}, that  melting of $J/\psi$ alone  cannot explain the data.   

The "Hydro+$J/\psi$" model has three parameters, the melting temperatures, $T_{J/\psi}$ and $T_\chi$ and the fraction $F$ of higher states ($\chi_c$+$\psi^\prime$).  If we include the cold nuclear matter effect, then $J/\psi$-nucleon absorption cross-section $\sigma_{abs}$,  can be considered as an additional parameter. However, as mentioned earlier, cold nuclear matter effect in d+Au collisions is  consistent with $\sigma_{abs}=2 \pm 1$ mb.
In the following, we assume that in Au+Au collisions also,  cold nuclear matter effect is also adequately explained in the Glauber model of nuclear absorption with $\sigma_{abs}$=2 mb.

Since the number of PHENIX data points are few, we do not fit simultaneously all the three parameters, $T_{J/\psi}$, $T_\chi$ and $F$. For fixed melting temperature $T_\chi$=1.-1.2$T_c$, we  vary the $T_{J/\psi}$ and $F$ to fit the PHENIX data. Note that in the present work $T_c$=164 MeV. 
The best-fitted values are shown in table 1.  Melting temperatures $T_{J/\psi}$ and $T_\chi$  are well determined, $T_{J/\psi}/T_c=2.08\pm 0.25$,
$T_\chi/T_c=1.1\pm 0.1$.   
The feed-down fraction however can only  be determined with large uncertainty, $F=0.32 \pm0.20$.  Quality of the PHENIX data is far from satisfactory, data points are few, and the associated error bars are large. If quality of the data is improved, the uncertainty in $F$ (and also in $T_{J/\psi}$ and $T_\chi$) can be reduced. The melting temperature $T_{J/\psi}$ and $T_\chi$ and also the feed-down fraction $F$  are in agreement with the results obtained by Gunji \etal \cite{Gunji:2007uy}. In Fig.\ref{F4}, the fit obtained to the data for $T_\chi$=$1.1T_c$, $T_{J/\psi}$=$2.08T_c$ and $F=0.32$ is shown. Sequential melting of charmonium states $\chi_c$, $\psi\prime$ and $J/\psi$ in an expanding QGP medium, well explain the PHENIX data. 
 
Sequential
melting of charmonium states in a deconfined medium can explain the PHENIX data even if we disregard the cold nuclear matter effect. 
 In table \ref{table1}, we have tabulated the best-fitted  melting temperature $T_{J/\psi}$ and the feed-down fraction $F$ obtained by fitting the data when    cold nuclear matter effect is neglected.
($\sigma_{abs}$=0). Whether the cold nuclear matter effect is included or not, within the error, the melting temperature $T_{J/\psi}$ remain same. However, the data require  higher feed-down fraction $F\approx 0.5$ for equivalent fit.   
  In Fig.\ref{F4}, we have shown the fit obtained to the data with $T_{J/\psi}=2T_c$, $T_\chi=1.1T_c$ and $F=0.5$. The fit is comparable to the one obtained with the cold nuclear matter effect included. 
The result is not unexpected.
In \cite{Chaudhuri:2007yy} it was shown that in the  phenomenological threshold model, sequential melting of charmonium states  can explain the PHENIX data, with or without the  cold nuclear matter effect. When cold nuclear matter effect is neglected for, the data demand higher feed-down fraction.  
Cold nuclear matter effects can be effectively mimicked by  melting of  higher states $\chi_c$'s and $\psi^\prime$.  The result underlies the importance of $J/\psi$ measurements in pA collisions. In heavy ion collisions, the feed-down fraction $F$ can  be  accurately estimated only when the cold nuclear  matter effect is accurately quantified in pA collisions.
 
The present analysis suggests that the centrality dependence of $J/\psi$ suppression in mid-rapidity Au+Au collisions at RHIC is consistent with sequential melting of $\chi_c$, $\psi\prime$ and $J/\psi$ in an expanding QGP medium. The melting temperature are well constrained, $T_{\chi_c}=T_{\psi\prime}=1-1.1T_c$,
$T_{J/\psi}=2-2.1T_c$. The feed-down fraction depends on whether the cold nuclear matter effect is included or not. Model require
$F=0.3 \pm 0.2$ if cold nuclear matter effect is quantified in the Glauber model with $J/\psi$-nucleon absorption cross-section $\sigma_{abs}$=2 mb. Larger fraction $F=0.5 \pm 0.20$ is required if cold nuclear matter effect is neglected.
 
In all the calculations presented here,
the decay width  for $J/\psi$   abruptly changes from $\infty$ to 0 at the melting temperature $T_{J/\psi}$ and similarly for the states $\chi_c$ and $\psi\prime$. 
Gunji \etal \cite{Gunji:2007uy} studied the effect of smoothening the decay width by using, 
 
 \begin{eqnarray} \nonumber
\Gamma_{dis}(T) &= & \infty; \hspace{1cm} T>T_{J/\psi} \\
\Gamma_{dis}(T) &= &\alpha (T/T_c -1)^2; \hspace{1cm}T<T_{J/\psi} \label{eq3}
\end{eqnarray}

\noindent where $\alpha$ is the thermal width of the state at $T/T_c$=2. NLO perturbative calculations suggest that
$\alpha > 0.4 GeV$ \cite{Park:2007zza}. However, for $\alpha \geq 0.4$, the PHENIX data are not well described. Data require that $\alpha \leq 0.1$, when smoothening effect is not large.

\begin{figure}[t]
\includegraphics[bb=40 287 532 769
 ,width=0.9\linewidth,clip]{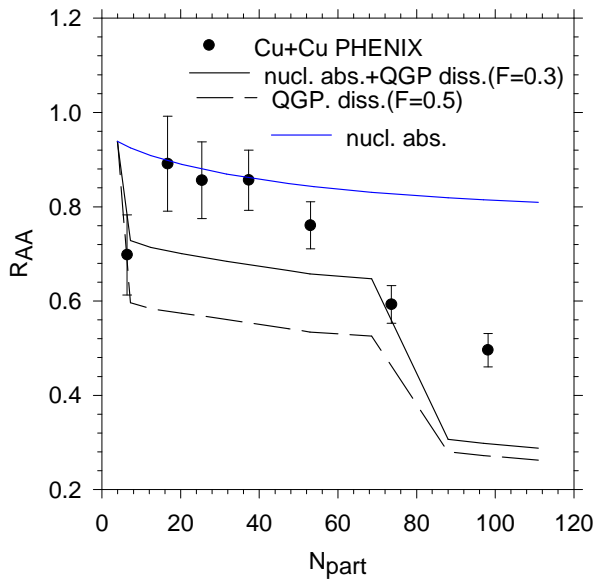}
\caption{(color online) The filled circles are PHENIX data on the centrality dependence of $J/\psi$ suppression in Cu+Cu collisions. The 
blue line is the absorption in Glauber model of nuclear absorption with $\sigma_{abs}$=2 mb. The solid    line is the suppression due to sequential melting of $\chi_c$, $\psi\prime$ and $J/\psi$ in an expanding QGP fluid, including the nuclear absorption. The dashed line is the same without the nuclear absorption.}
 \label{F5}
\end{figure}

\subsection{$J/\psi$ suppression in Cu+Cu collisions}

Recently PHENIX collaboration published their analysis of $J/\psi$ measurements in Cu+Cu collisions \cite{Adare:2008sh}. 
In Fig.\ref{F5}, PHENIX data \cite{Adare:2008sh} on the centrality dependence of $J/\psi$ are shown.
In the most central collision, $J/\psi$'s are suppressed by a factor of $\sim$ 2. Survival probability of $J/\psi$ increases as the collision centrality decreases till $N_{part}$=16.7. In more peripheral ($N_{part}$=6.4), collisions, survival probability decreases again. The  decrease in survival probability as
participant number decreases from $N_{part}$=16.7 to 6.4 
is interesting. All the theoretical models predict continuous
increase of suppression as the collision centrality increases.
It was also noted \cite{Adare:2008sh} that at comparable participant number, $J/\psi$ are suppressed similarly in Au+Au and in Cu+Cu collisions. However, there is a major difference between $J/\psi$ suppression in Cu+Cu and in Au+Au collisions. 
In Fig.\ref{F5}, the blue line is the suppression in Cu+Cu collisions, calculated in a Glauber model of nuclear absorption, with $\sigma_{abs}$=2 mb. As noted by the PHENIX collaboration, suppression in peripheral collisions (excluding $N_{part}$=6.4) is consistent with the nuclear absorption alone. 
In contrast, $J/\psi$ suppression in peripheral Au+Au collisions are not consistent with Glauber model of nuclear absorption.

\begin{figure}[t]
\includegraphics[bb=38 292 527 769
 ,width=0.9\linewidth,clip]{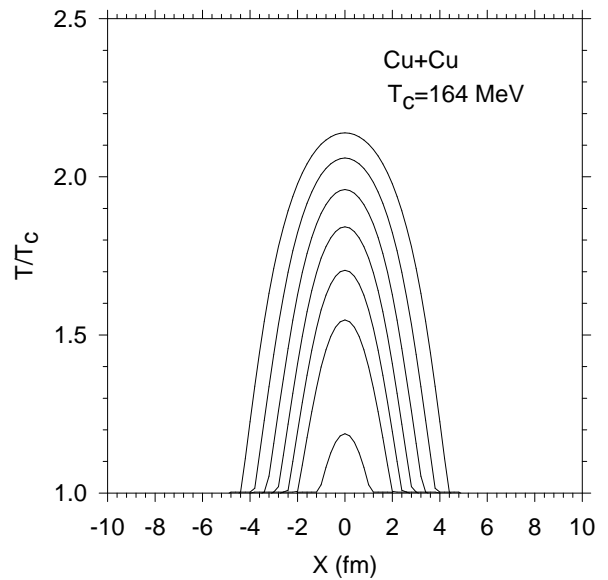}
\caption{Initial temperature $T(x,y=0)$, 
in unit of the critical temperature, of the QGP fluid in Cu+Cu collisions at impact parameter b= 8.83, 7.55,6.82,6.00,5.05,3.87 and 2.29 fm (bottom to top), respectively.}
 \label{F6}
\end{figure}

For hydrodynamical evolution of QGP fluid in Cu+Cu collisions,
we use Eq.\ref{eq1} to calculate the initial energy density
in the transverse plane. The constant $\varepsilon_0$ in Eq.\ref{eq1}, depend only on the collision energy, not on centrality of the collisions. At the initial time $\tau_i$=0.6 fm/c, the initial energy density  is $\varepsilon\approx $30 $GeV/fm^3$ or initial central entropy density is $S_{ini}=110 fm^{-3}$.   
In Fig. \ref{F6},  the initial temperatures $T(x,y=0)$ of the fluid in b=2.29, 3.87, 5.05, 6.0, 6.82, 7.55 and 8.83 fm Cu+Cu collisions are shown. The impact parameters roughly corresponds to   participant numbers $N_{part}$=98.2, 73.6, 53.0, 37.3, 25.4, 16.7 and 6.4 respectively.     
In all these collisions, peak temperature of the fluid exceeds the melting temperature $T_\chi=T_{\chi_c}=T_{\psi\prime}$=$1-1.1Tc$. 
Consequently, in all the centrality ranges of collisions, melting of the states $\chi_c$ and $\psi^\prime$ will contribute to the  
$J/\psi$ suppression. Immediately we find that $J/\psi$ suppression in peripheral Cu+Cu collisions is not consistent with sequential melting of charmonium states. Peripheral collisions are consistent with cold nuclear matter effect and cannot accommodate additional suppression due to higher states $\chi$ and $\psi\prime$.  
 
It is clearly shown in Fig.\ref{F5}.
When cold nuclear matter effect is included,
PHENIX Au+Au data are explained (Fig.\ref{F4}) with  melting temperatures ,  $T_{J/\psi}=2.08 T_c$, $T_\chi=1.1T_c$ and feed-down fraction   $F=0.3$. In Fig.\ref{F5}, the solid line is the centrality dependence of $J/\psi$ suppression in Cu+Cu collisions, with cold nuclear matter effect included. The parameter values are unchanged, i.e.
$T_{J/\psi}=2.08 T_c$, $T_\chi=1.1T_c$ and    $F=0.3$.
As expected, the Hydro+$J/\psi$ model produces more suppression than required by the data.  

Will the data will be explained without any nuclear absorption
but with increased feed-down fraction?
As shown earlier, the PHENIX Au+Au data on $J/\psi$ suppression are explained with same melting temperatures but with increased feed-down fraction, $F\approx 0.5$.
In Fig.\ref{F5}, the dashed line is the centrality dependence of $J/\psi$ suppression without any nuclear absorption, but with increased feed-down fraction $F=0.5$. The data are not explained either. In contrast to Au+Au collisions, sequential melting of charmonium states  in an expanding QGP medium do not explain the centrality dependence of $J/\psi$ suppression in Cu+Cu collisions.
It may be noted that the present model neglect the recombination of $c\bar{c}$ pairs. Inclusion of recombination effect will reduce $J/\psi$ suppression.
A shown earlier, within the present model, $J/\psi$ suppression in Au+Au collisions do not require any recombination.
It is unlikely that recombination effect will be important in Cu+Cu collisions but not in Au+Au collisions.  It appears that, mechanism of $J/\psi$ suppression in Au+Au and in Cu+Cu collisions is different.

\begin{figure}[ht]
\includegraphics[bb=55 291 544 770
,width=0.9\linewidth,clip]{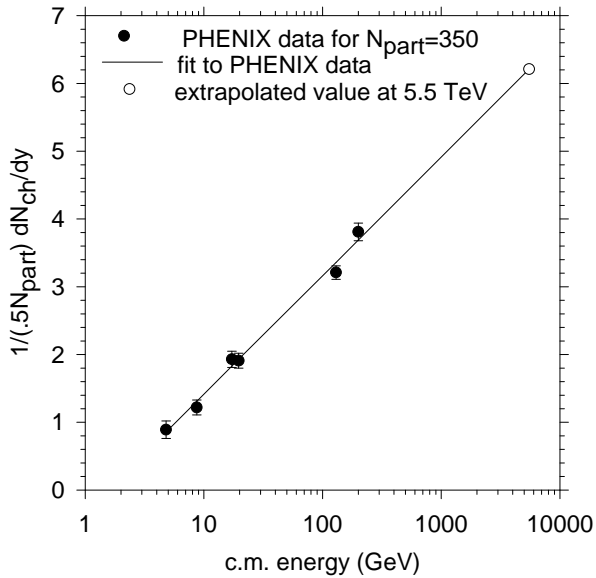}
\caption{Filled circles are the PHENIX data for the charged particle multiplicity per participant pair $\frac{1}{.5N_{part}}\frac{dN}{dy}$ as a function of c.m. energy for participant number $N_{part}$=350. The solid line is a fit to the 
PHENIX data by Eq.\ref{eq3}. The unfilled circle is the extrapolated value of $\frac{1}{.5N_{part}}\frac{dN}{dy}$    at LHC energy $\sqrt{s}$=5.5 TeV, for participant number $N_{part}=350$.}
\label{F7}
\end{figure} 

\section{$J/\psi$ suppression in Pb+Pb collision at LHC}

At the Large Hadron Collider (LHC), it is planned to collide Pb ions at centre of mass energy  $\sqrt{s}$=5500GeV. Signals of the deconfinement phase transition are expected to  be better defined at LHC than at RHIC collisions. One wonders whether in LHC energy collisions, $J/\psi$ will be more suppressed than at RHIC energy? Recently,  in the phenomenological threshold model,  it was predicted that suppression pattern in Pb+Pb collisions at LHC energy will be similar to that in Au+Au collisions at RHIC \cite{Chaudhuri:2007yy}. The reason is understood.
In the threshold model, suppression depends on the local transverse density. If the local transverse density exceed a critical value, $J/\psi$'s are suppressed. The local transverse density 
depend marginally on the collisions energy and remain essentially same at RHIC and LHC energy collisions. In the "Hydro+$J/\psi$" model, $J/\psi$ suppression depend directly on the "local fluid temperature" and in LHC energy collisions, local temperature will be substantially greater than that in RHIC energy collisions.  One expects enhanced suppression at LHC energy.

\begin{figure}[ht]
\includegraphics[bb=38 287 532 769
 ,width=0.9\linewidth,clip]{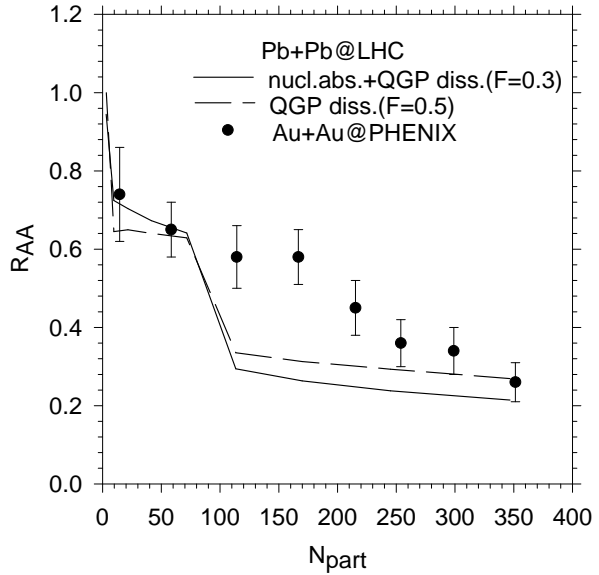}
\caption{ Hydro+$J/\psi$ model predictions for the centrality dependence of $J/\psi$ suppression in Pb+Pb collisions at LHC energy. The melting temperature for the states $J/\psi$,  $\psi\prime$ and $\chi_c$ are $T_{J/\psi}=2.1 T_c$ and $T_\chi=T_{\chi_c}=T_{\psi\prime}=1.1T_c$
The solid line is obtained with feed-down fraction $F=0.3$  and 
including the nuclear absorption effect with $\sigma_{abs}$=2 mb. The dashed line is without the nuclear absorption effect but with increased feed-down fraction F=0.5. For comparison,
PHENIX data on the centrality dependence of
 $J/\psi$ suppression in Au+Au collisions are also shown.}
\label{F8}
\end{figure}

For predicting $J/\psi$ suppression in Pb+Pb collisions at LHC, we have to guess the initial energy density distribution of the QGP fluid. PHENIX collaboration \cite{Adler:2004zn} has tabulated the average charged particle multiplicity as
a function of collision energy for a range of collision centrality. 
In Fig.\ref{F7}, for participant number $N_{part}$=350, the average multiplicity $\frac{1}{.5N_{part}} \frac{dN_{ch}}{d\eta}$ is shown as a function of collision energy . The multiplicity increases logarithmically with energy,

\begin{equation} \label{eq4}
\frac{dN_{ch}}{d\eta}=A+ B \ln \sqrt{s},
\end{equation}

\noindent with $A=-0.33$ and $B=0.75$. We use the relation to extrapolate to LHC energy $\sqrt{s}$=5.5 TeV. The extrapolated value of average charged particle multiplicity in LHC energy is $\sim 927\pm 70$.   We adjust
the central entropy density to $S_{ini}$=180 $fm^{-3}$ such that a $N_{part}$=350 Pb+Pb collisions produce $\sim$ 900 charged particles. The participant numbers in Pb+Pb collisions are calculated with NN inelastic cross-section $\sigma_{inel}$=70 mb.
Entropy density $S_{ini}$=180 $fm^{-3}$ corresponds to central temperature corresponds to $T_i$=421 MeV.
This can be contrasted to central temperature $T_i$=357 MeV in  Au+Au collisions . Compared to Au+Au collisions at RHIC,  in Pb+Pb collisions at LHC, central temperature approximately $\sim$ 20\% higher.  

\begin{figure}[ht]
\includegraphics[bb=25 292 582 769
 ,width=0.9\linewidth,clip]{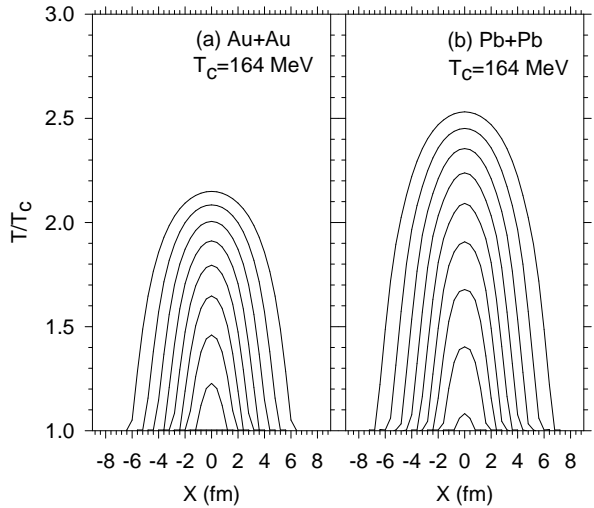}
\caption{In panel (a) and (b), initial temperature $T(x,y=0)$ of the QGP fluid, in unit of the critical temperature ($T_c$),  in 
Au+Au collisions at RHIC and in Pb+Pb collisions at LHC, are compared. The lines (top to bottom) corresponds to
0-10\%, 10-20\%, 20-30\%, 30-40\%, 40-50\%, 50-60\%, 60-70\%, 70-80\% collisions.}
 \label{F9}
\end{figure}

In Fig.\ref{F8},   Hydro+$J/\psi$ model predictions for
the centrality dependence of $J/\psi$ suppression in Pb+Pb collisions are shown. The melting temperature of the states $J/\psi$ and $\chi_c$ and $\psi\prime$ are property of the deconfined medium and do not depend on the collision energy. We use the values extracted from the analysis of RHIC data, $T_{J/\psi}=2-2.1T_c$ and $T_\chi=1-1.1T_c$. In Fig.\ref{F8}, the solid line is the suppression obtained with 
$T_{J/\psi}=2.1T_c$ and $T_\chi=1.1T_c$, $F=0.3$, the cold nuclear matter effect is included. Nearly similar suppression is obtained when cold nuclear matter effect is neglected but, the feed-down fraction is increased to $F=0.5$ (the dashed line in Fig.\ref{F8}).
  For comparison, we have shown the 
PHENIX data on $J/\psi$ suppression in Au+Au collisions. In very central collisions or very peripheral collisions, $J/\psi$ are suppressed similarly in RHIC and LHC energy. But in mid central collisions, $J/\psi$'s are more suppressed in LHC energy than in RHIC energy collisions. One also notices that while in Au+Au collisions at RHIC, the  change in slope occur around $N_{part}\approx$ 150, in Pb+Pb collisions at LHC, the change occur at lower participant number $N_{part}\approx$ 70. 
The reason can be understood from Fig. \ref{F9},  
 where we have compared initial temperatures $T(x,y=0)$
in 0-10\%, 10-20\%, 20-30\%, 30-40\%, 40-50\%, 50-60\%, 60-70\%, 70-80\% and 80-90\% in Au+Au collisions RHIC and in Pb+Pb collisions at LHC. Both in Au+Au and Pb+Pb collisions, peak temperature in 0-10\% centrality collisions exceed the melting temperature $T_{J/\psi}= 2.1T_c$. In 0-10\% centrality collisions, $J/\psi$ will be   suppressed both in RHIC and LHC energy collisions. 
However, it is not so in less central collisions,  e.g.
in 20-30\% centrality collisions, while peak temperature exceed the melting temperature $T_{J/\psi}$ in Pb+Pb collisions at LHC, in Au+Au collisions, the peak temperature is less than the melting temperature. $J/\psi$ will survive in 20-30\% centrality Au+Au collision at RHIC but they will be melted in 20-30\% centrality Pb+Pb collisions at LHC.  
Thus $J/\psi$'s can survive in mid-central Au+Au collisions at RHIC, but are suppressed in mid-central Pb+Pb collisions. 
For the very reason, the change of slope in the suppression pattern also occurs at lower participant number. Compared to RHIC energy collisions, the melting temperature $T_{J/\psi}$ is reached in lower participant number collisions at LHC energy.


\section{Summary and conclusions} 

To summarise, we have developed a "Hydro+$J/\psi$" model to study $J/\psi$ suppression in an expanding Quark-Gluon-Plasma.  The space-time evolution of the QGP fluid is obtained by solving the 
hydrodynamic equations for ideal fluid in 2+1 dimensions. At the initial time,
  $J/\psi$'s are randomly distributed in the transverse plane. As the fluid evolve in time, the free streaming $J/\psi$'s are completely suppressed if the local fluid  temperature exceed a critical temperature $T_{J/\psi}$. Similarly, the states $\chi_c$ and $\psi\prime$ are assumed to melt above a critical temperature $T_\chi$. The melting temperatures $T_{J/\psi}$ and $T_\chi$ and the feed-down fraction $F$ from the higher states $\chi_c$ and $\psi\prime$, are fitted to reproduce the PHENIX data on the centrality dependence of $J/\psi$ suppression in Au+Au collisions at mid-rapidity. The PHENIX data are well explained with $T_{J/\psi}\approx 2-2.1 T_c$, $T_\chi\approx 1-1.1 T_c$. The feed-down fraction depends on whether or not  $J/\psi$ suppression in cold nuclear matter is included. If cold nuclear matter effect is included, data require $F \approx0.3$. Fraction $F$ increases to
$F\approx $0.5, if the cold nuclear  matter effect is neglected. It appears that, to a large extent, suppression of the states $\chi_c$ and $\psi\prime$,  can mimic the cold nuclear matter effect. While sequential melting of $\chi_c$, $\psi\prime$ and $J/\psi$ in an expanding QGP fluid well explain the $J/\psi$ suppression in Au+Au collisions, the model fails to reproduce the experimental data in Cu+Cu collisions. In Cu+Cu collisions, "Hydro+$J/\psi$" model produces more suppression than required by the data, indicating   that $J/\psi$  suppression mechanism in Au+Au and in Cu+Cu collisions are different. We have also given prediction for the centrality dependence of $J/\psi$ suppression in Pb+Pb collisions at LHC energy. The model predicts more suppression in LHC than in RHIC energy collisions.

 \begin{widetext}
\begin{table}[ht] 
\caption{Melting temperatures (in unit of $T_c$)  of direct $J/\psi$, ($T_{J/\psi}$),
the states $\chi_c$ and $\psi^\prime$ ($T_\chi$) and the fraction of the higher states $\chi_c$ and $\psi^\prime$. For fixed melting temperature $T_\chi$, the PHENIX data are fitted by varying the melting temperature $T_{J/\psi}$ and $F$,  Values obtained with ($\sigma_{abs}$=2 mb) and without ($\sigma_{abs}$= 0 mb) the cold nuclear matter effects are shown.}
  \begin{tabular}{|c|ccc||ccc|}\hline
 &  & $\sigma_{abs}$=2 mb & &  & $\sigma_{abs}$=0 &   \\ 
$\frac{T_\chi}{T_c}$ & $\frac{T_{J/\psi}}{T_c}$ & $F$ & $\chi^2/d.o.f$ & $\frac{T_{J/\psi}}{T_c}$  & F & $\chi^2/F$ \\ \hline
1.00 & $2.08\pm 0.18$ & $0.30\pm 0.17$ & 0.72 &$2.07\pm 0.22$ & $0.49\pm 0.22$ & 1.50 \\ \hline
1.05 & $2.08\pm 0.23$ & $0.31\pm 0.18$ & 0.72 &$2.00\pm 0.22$ & $0.51\pm 0.18$ & 0.89 \\ \hline
1.10 & $2.08\pm 0.13$ & $0.32\pm 0.18$ & 0.68  & $2.00\pm 0.33$ &  $0.51\pm 0.18$ & 0.90\\ \hline
1.15 & $2.07\pm 0.46$ & $0.32\pm 0.23$ & 0.97  & $2.05\pm 0.37$ &  $0.55\pm 0.25$ & 1.48\\ \hline
1.20 & $2.08\pm 0.27$ & $0.32\pm 0.23$ & 0.96 & $2.00\pm 0.51$ &
$0.56 \pm 0.26$ & 1.47 \\ \hline
  \end{tabular} \label{table1}
\end{table}
\end{widetext}


\end{document}